\documentstyle[11pt]{article}
\begin{document}
\title{Self-Duality beyond Chiral {\em p}-Form Actions}
\author{{Yan-Gang Miao${}^{{\rm a,b},1,2}$, R. Manvelyan${}^{{\rm a},1,3}$ and H.J.W. M$\ddot{\rm u}$ller-Kirsten
${}^{{\rm a},4}$}\\
{\small ${}^{\rm a}$ Department of Physics, University of Kaiserslautern,
P.O. Box 3049,}\\
{\small D-67653 Kaiserslautern, Germany}\\
{\small ${}^{\rm b}$ Department of Physics, Xiamen University, Xiamen 361005,}\\
{\small People's Republic of China}}
\date{}
\maketitle
\footnotetext[1]{Alexander von Humboldt Fellow.}
\footnotetext[2]{E-mail: miao@physik.uni-kl.de}
\footnotetext[3]{E-mail: manvel@physik.uni-kl.de. On leave from Yerevan Physics Institute.}
\footnotetext[4]{E-mail: mueller1@physik.uni-kl.de}
\vskip 48pt
\begin{center}{\bf Abstract}\end{center}
\baselineskip 22pt
\par
The self-duality of chiral {\em p}-forms was originally investigated by Pasti, Sorokin and Tonin in a manifestly Lorentz covariant action with non-polynomial auxiliary fields. The investigation was then extended to other chiral {\em p}-form actions. In this paper we point out that the self-duality appears in a wider context of theoretical models that relate to chiral {\em p}-forms. We demonstrate this by considering the interacting model of Floreanini-Jackiw chiral bosons and gauge fields, the generalized chiral Schwinger model (GCSM) and the latter's gauge invariant formulation, and discover that the self-duality of the GCSM corresponds to the vector and axial vector current duality.
\vskip 24pt
PACS numbers: 11.10.-z; 11.15.-q; 11.30.-j
\newpage
\section{Introduction}
\par
    Chiral {\em p}-forms have attracted much attention because they play an important
role in many theoretical models. In $D=2$ dimensional space-time, chiral bosons ($p=0$) occur as
basic ingredients and elements in the formulation of heterotic strings [1] and
in a number of statistical systems [2]. In $D>2$ dimensional space-time, chiral 2- and 4-forms are relevant to the M-theory five-brane [3,4]
 and type
IIB supergravity [5-7], respectively.

Chiral {\em p}-forms are described by an
antisymmetric {\em p}th order tensor $A^{(p)}$ in the $D=2(p+1)$ dimensional
space-time, whose
external differential $F^{(p+1)}(A)=dA^{(p)}$ satisfies the self-duality
condition
\begin{equation}
{\cal F}^{(p+1)} \equiv F^{(p+1)}(A)-{}^{\ast}F^{(p+1)}(A)=0,
\end{equation}
where ${}^{\ast}F^{(p+1)}(A)$ is
defined as the dual
partner of $F^{(p+1)}(A)$. In the
space with the
Lorentzian metric signature, the
self-duality requires $A^{(p)}$ to be
real if {\em p} is
even, or complex if {\em p} is odd. In the latter case the theory can 
equivalently be described by a pair of real antisymmetric tensor fields related
by a duality condition.

Since the equation of motion of a chiral {\em p}-form, i.e., the
self-duality condition, is first order with respect to the derivatives of
space and time, it is a key problem to construct the corresponding action and
then to quantize the theory consistently. To this end, various formulations of
actions have been proposed [8-14]. These actions can be classified by
manifestly Lorentz covariant versions [8-12] and non-manifestly Lorentz
covariant versions [13,14] when one emphasizes their formalism under the
Lorentz transformation, or by polynomial versions [8-11] and non-polynomial
version [12] when one focuses on auxiliary fields introduced in the actions.
Incidentally, there are no auxiliary fields introduced in the non-manifestly
Lorentz covariant actions [13,14].

It is noticeable that these chiral {\em p}-form actions have close relationships among one another. The recently constructed  Pasti-Sorokin-Tonin action [12]
reduces to the
non-manifestly covariant Floreanini-Jackiw one [13] provided
appropriate gauge fixing conditions are chosen. On the other hand, it 
turns into the
McClain-Wu-Yu formulation [11] if one gets rid of the Pasti-Sorokin-Tonin action's non-polynomiality and
eliminates its scalar auxiliary field at the price of introducing auxiliary
({\em p}+1)-forms, or, vice versa, if one consistently truncates the McClain-Wu-Yu action's
infinite tail and puts on its end the auxiliary scalar field. Moreover, it has been shown [15] that the Pasti-Sorokin-Tonin action follows directly from the Kavalov-Mkrtchyan formulation [9] that is the Siegel action's generalization with an auxiliary higher (than two) rank tensor field. 

In our previous work [16], the duality symmetries of four chiral {\em p}-form actions are investigated. We discover that the Siegel, Floreanini-Jackiw and Pasti-Sorokin-Tonin actions have self-duality with respect to a common anti-dualization of chiral boson ($p=0$) and chiral 2-form fields, respectively, while the Srivastava action is self-dual with respect to a generalized dualization of the corresponding chiral fields. The result can be extended to the general case, that is, the self-duality remains in $D=2(p+1)$ space-time dimensions. In addition, we note here that the Kavalov-Mkrtchyan formulation [9] also has self-duality with respect to an anti-dualization of chiral 2-form fields, which can be obtained straightforwardly along the line of Ref.[16]. 

The self-duality of chiral {\em p}-forms was first investigated [12] in the Pasti-Sorokin-Tonin action and then extended [16] to others. Here we point out that the self-duality appears in a wider context of theoretical models that relate to chiral {\em p}-forms. As examples, we choose the interacting model of Floreanini-Jackiw chiral bosons and gauge fields [17], the generalized chiral Schwinger model (GCSM) [18] and its gauge invariant formulation [19]. These models are usually dealt with as a `theoretical laboratory' in illustrating new aspects of field theory and have been utilized in a large amount of literature.

    The paper is arranged as follows. In Sects. 2, 3
and 4, we discuss the duality symmetries of the three models 
one by one 
and finally make
a conclusion in Sect.5.

    The notation we use throughout this paper is
\begin{eqnarray}
g_{00}=-g_{11}=1, {\epsilon}^{01}=-{\epsilon}_{01}=1,\nonumber \\
{\gamma}^0={\sigma}_1, {\gamma}^1=-i{\sigma}_2,  {\gamma}^5={\sigma}_3,\nonumber \\
\Box={\partial}_{\mu}{\partial}^{\mu},  \dot{\phi}={\partial}_{0}{\phi},  {\phi}^{\prime}={\partial}_{1}{\phi},
\end{eqnarray}
${\sigma}_i$ being the Pauli matrices.
\section{Self-duality of the interacting model of Floreanini-Jackiw chiral bosons and gauge fields}
\par
    We begin with the action of this interacting theory [17]
\begin{eqnarray}
S&=&\int d^{2}x\left[\dot{\phi}{\phi}^{\prime}
-({\phi}^{\prime})^2+2e{\phi}^{\prime}(A_{0}-A_{1})\right.\nonumber
\\
& &\left.-{\frac 1 2}e^2(A_{0}-A_{1})^2+{\frac 1 2}
e^2aA_{\mu}A^{\mu} -{\frac 1 4}F_{{\mu}{\nu}}F^{{\mu}{\nu}}\right],
\end{eqnarray}
where $\phi$ is a scalar field, $A_{\mu}$ a gauge field and $F_{{\mu}{\nu}}$
its field strength;
{\em e} is the electric charge and {\em a} a real parameter caused by
ambiguity in
bosonization. It is a non-manifestly Lorentz covariant action but indeed has
Lorentz invariance, and the spectrum includes one massive free scalar boson and one free chiral boson [17]. In the following, 
the first three terms in eq.(3) are important, while the last three that
relate only to gauge fields have nothing to do with the duality property
of the action.

    By introducing two auxiliary vector fields $F_{\mu}$ and $G^{\mu}$, we
construct a new action to replace eq.(3)
\begin{eqnarray}
S=\int d^{2}x\left[F_{0}F_{1}
-(F_{1})^2+2eF_{1}(A_{0}-A_{1})
-{\frac 1 2}e^2(A_{0}-A_{1})^2\right.\nonumber \\
\left.+{\frac 1 2}e^2aA_{\mu}A^{\mu}
-{\frac 1 4}F_{{\mu}{\nu}}F^{{\mu}{\nu}}
+G^{\mu}(F_{\mu}-{\partial}_{\mu}{\phi})\right],
\end{eqnarray}
where $F_{\mu}$ and $G^{\mu}$ are treated as independent fields. Variation of
eq.(4) with respect to the Lagrange multiplier $G^{\mu}$ gives
$F_{\mu}={\partial}_{\mu}{\phi}$,
which yields the equivalence between the two actions eqs.(3) and (4).
Furthermore, variation of eq.(4) with respect to $F_{\mu}$ leads to the
expression of $G^{\mu}$ in terms of $F_{\mu}$
\begin{eqnarray}
G^{0}&=&-F_{1},\nonumber \\
G^{1}&=&-F_{0}+2F_{1}-2e(A_{0}-A_{1}).
\end{eqnarray}
Solving these for $F_{\mu}$
\begin{eqnarray}
F_{0}&=&-2G^{0}-G^{1}-2e(A_{0}-A_{1}),\nonumber \\
F_{1}&=&-G^{0}.
\end{eqnarray}
If we define ${\cal F}_{\mu}=F_{\mu}-{\epsilon}_{{\mu}{\nu}}F^{\nu}$ and
${\cal G}_{\mu}=G_{\mu}-{\epsilon}_{{\mu}{\nu}}G^{\nu}$,
we find that they satisfy the relation
\begin{equation}
{\cal F}_{\mu}=-{\cal G}_{\mu}+2e({\epsilon}_{{\mu}{\nu}}-g_{{\mu}{\nu}})
A^{\nu},
\end{equation}
which is different from that of the free Floreanini-Jackiw case because of
interactions. In other words, if the interaction did not exist, i.e., $e=0$,
eq.(7) would reduce to the free theory case
${\cal F}_{\mu}=-{\cal G}_{\mu}$ [16].
Substituting eq.(6) into eq.(4), we obtain the dual action in terms of
$G^{\mu}$
\begin{eqnarray}
S_{dual}&=&\int d^{2}x\left[-(G^{0})^2
-G^{0}G^{1}-2eG^{0}(A_{0}-A_{1})\right.\nonumber \\
& &\left.-{\frac 1 2}e^2(A_{0}-A_{1})^2
+{\frac 1 2}e^2aA_{\mu}A^{\mu}
-{\frac 1 4}F_{{\mu}{\nu}}F^{{\mu}{\nu}}
+{\phi}{\partial}_{\mu}G^{\mu}\right].
\end{eqnarray}
Variation of eq.(8) with respect to $\phi$ gives ${\partial}_{\mu}G^{\mu}=0$,
whose solution should be
\begin{equation}
G^{\mu}(\rho)=-{\epsilon}^{{\mu}{\nu}}{\partial}_{\nu}{\rho}
\equiv -{\epsilon}^{{\mu}{\nu}}F_{\nu}({\rho}),
\end{equation}
where ${\rho}(x)$ is an arbitrary scalar field. Substituting eq.(9) into
eq.(8), we obtain the dual action in terms of $\rho$
\begin{eqnarray}
S_{dual}&=&\int d^{2}x\left[\dot{\rho}{\rho}^{\prime}
-({\rho}^{\prime})^2+2e{\rho}^{\prime}(A_{0}-A_{1})\right.\nonumber
\\
& &\left.-{\frac 1 2}e^2(A_{0}-A_{1})^2+{\frac 1 2}e^2aA_{\mu}A^{\mu}
-{\frac 1 4}F_{{\mu}{\nu}}F^{{\mu}{\nu}}\right].
\end{eqnarray}
This action has the same formulation as the original action eq.(3) only with the
replacement of $\phi$ by $\rho$. Note that because of interactions, ${\phi}(x)$
no longer coincides with ${\rho}(x)$ up to a constant on the mass shell, which
is
different from that of the free theory case [16]. This means that eq.(9) shows a
generalized anti-dualization of $F_{\mu}(\phi)$ and $G_{\mu}(\rho)$. We emphasize that the mass shell, i.e., the self-duality condition, is not necessary for self-duality of actions because it can not directly be imposed on actions. Therefore, we prove
that the interacting model of Floreanini-Jackiw chiral bosons and gauge fields has self-duality
with respect to the generalized anti-dualization of `field strength'
expressed by eq.(9). Incidentally, if we choose the solution
$G^{\mu}(\rho)={\epsilon}^{{\mu}{\nu}}{\partial}_{\nu}{\rho}$
instead of eq.(9), the dual action has a minus sign in the third term.
However, the spectrum is the same
whether the third term of eq.(10) is positive or negative. As a result, the self-duality remains with respect to the generalized (anti-)dualization of $G^{\mu}{(\rho)}$ and $F_{\mu}{(\phi)}$. 
\section{Self-duality of the generalized chiral Schwinger model}
\par
The fermionic action of the GCSM takes the form [18]
\begin{equation}
S_{F}=\int d^{2}x\left\{\bar{\psi}{\gamma}^{\mu}[i{\partial}_{\mu}+e\sqrt{\pi}(1+r{\gamma}^{5})A_{\mu}]{\psi}-{\frac 1 4}F_{{\mu}{\nu}}F^{{\mu}{\nu}}\right\},
\end{equation}
where $\psi$ is a massless spinor, $A_{\mu}$ a gauge field and $F_{{\mu}{\nu}}$ its field strength. The quantity {\em r} is a real parameter interpolating between the vector ($r=0$) and the chiral ($r=\pm 1$) Schwinger models. Since a bosonic action presents an anomaly at tree level while a fermionic action does at least at one-loop level, we prefer the bosonic version which can be obtained by the operatorial [20] or the path-integral bosonization [21]. The bosonic action can be written as follows [18]
\begin{equation}
S_{B}=\int d^{2}x\left[{\frac 1 2}({\partial}_{\mu}\phi)({\partial}^{\mu}\phi)+eA^{\mu}({\epsilon}_{{\mu}{\nu}}-rg_{{\mu}{\nu}}){\partial}^{\nu}{\phi}
+{\frac 1 2}e^{2}aA_{\mu}A^{\mu}-{\frac 1 4}F_{{\mu}{\nu}}F^{{\mu}{\nu}}\right],
\end{equation}
where $\phi$ is an auxiliary scalar field introduced in order to result in a local $S_{B}$, and {\em a} is a real parameter which expresses the ambiguity in the bosonization procedure. From the derivation of the bosonic action, it is clear that $S_{B}$ is equivalent to $S_{F}$ in the sense that both actions lead to the same generating functional
\begin{eqnarray}
Z[A]&=&\int d{\psi}d\bar{\psi}{\rm exp}(iS_{F})=\int d{\phi}{\rm exp}(iS_{B})\nonumber \\
&=&{\rm exp}\left\{i\int d^{2}x\left[{\frac 1 2}e^{2}A^{\mu}({\epsilon}_{{\mu}{\alpha}}-rg_{{\mu}{\alpha}})\frac{{\partial}^{\alpha}{\partial}^{\beta}}{\Box}({\epsilon}_{{\beta}{\nu}}+rg_{{\beta}{\nu}})A^{\nu}\right.\right.\nonumber \\
& &\left.\left.+{\frac 1 2}e^{2}aA_{\mu}A^{\mu}-{\frac 1 4}F_{{\mu}{\nu}}F^{{\mu}{\nu}}\right]\right\},
\end{eqnarray}
where a field-independent constant has been dropped in the last equality.

In order to discuss the duality of the GCSM action, we introduce two vector fields $F_{\mu}$ and $G^{\mu}$, and replace eq.(12) by the following action
\begin{eqnarray}
S&=&\int d^{2}x\left[{\frac 1 2}F_{\mu}F^{\mu}+eA^{\mu}({\epsilon}_{{\mu}{\nu}}-rg_{{\mu}{\nu}})F^{\nu}\right.\nonumber \\
& &\left.+{\frac 1 2}e^{2}aA_{\mu}A^{\mu}-{\frac 1 4}F_{{\mu}{\nu}}F^{{\mu}{\nu}}+G^{\mu}(F_{\mu}-{\partial}_{\mu}\phi)\right],
\end{eqnarray}
where $F_{\mu}$ and $G^{\mu}$ act, at present, as independent auxiliary fields. Variation of eq.(14) with respect to $G^{\mu}$ gives $F_{\mu}={\partial}_{\mu}\phi$, which yields the classical equivalence between actions eqs.(12) and (14). On the other hand, variation of eq.(14) with respect to $F_{\mu}$ leads to the relation of $F_{\mu}$ and $G^{\mu}$
\begin{equation}
F_{\mu}=-G_{\mu}+e({\epsilon}_{{\mu}{\nu}}+rg_{{\mu}{\nu}})A^{\nu}.
\end{equation}
Substituting eq.(15) into the action eq.(14), we obtain the dual action of the GCSM
\begin{eqnarray}
S_{dual}&=&\int d^{2}x\left[-{\frac 1 2}G_{\mu}G^{\mu}-eA^{\mu}({\epsilon}_{{\mu}{\nu}}-rg_{{\mu}{\nu}})G^{\nu}\right.\nonumber \\
& &\left.+{\frac 1 2}e^{2}(a+1-r^{2})A_{\mu}A^{\mu}-{\frac 1 4}F_{{\mu}{\nu}}F^{{\mu}{\nu}}+{\phi}{\partial}_{\mu}G^{\mu}\right].
\end{eqnarray} 
Variation of eq.(16) with respect to $\phi$ gives ${\partial}_{\mu}G^{\mu}=0$, whose solution should be
\begin{equation}
G^{\mu}(\rho)={\pm}{\epsilon}^{{\mu}{\nu}}{\partial}_{\nu}{\rho}{\equiv}{\pm}{\epsilon}^{{\mu}{\nu}}F_{\nu}(\rho),
\end{equation}
where ${\rho}(x)$ is an arbitrary scalar field. When eq.(17) is substituted into eq.(16), we get the dual action in terms of $\rho$
\begin{eqnarray}
S_{dual}=\int d^{2}x\left[{\frac 1 2}({\partial}_{\mu}\rho)({\partial}^{\mu}\rho){\pm}eA^{\mu}(r{\epsilon}_{{\mu}{\nu}}-g_{{\mu}{\nu}}){\partial}^{\nu}{\rho}\right.\nonumber \\
\left.+{\frac 1 2}e^{2}(a+1-r^{2})A_{\mu}A^{\mu}-{\frac 1 4}F_{{\mu}{\nu}}F^{{\mu}{\nu}}\right].
\end{eqnarray}

In order to make a comparison between the action eq.(12) and its dual partner eq.(18), we first introduce three new parameters $r^{\prime}$, $e^{\prime}$ and $a^{\prime}$ defined by
\begin{equation}
r^{\prime}=\frac{1}{r}, e^{\prime}={\pm}er, a^{\prime}=\frac{a+1-r^{2}}{r^2},
\end{equation}
where $r\not=0$ in general, and rewrite eq.(18) as
\begin{equation}
S_{dual}=\int d^{2}x\left[{\frac 1 2}({\partial}_{\mu}{\rho})({\partial}^{\mu}{\rho})+e^{\prime}A^{\mu}({\epsilon}_{{\mu}{\nu}}-r^{\prime}g_{{\mu}{\nu}}){\partial}^{\nu}{\rho}
+{\frac 1 2}{e^{\prime}}^{2}a^{\prime}A_{\mu}A^{\mu}-{\frac 1 4}F_{{\mu}{\nu}}F^{{\mu}{\nu}}\right],
\end{equation}
which has the same form as eq.(12) with the replacements of $\phi$, {\em r}, {\em e} and {\em a} by $\rho$, $r^{\prime}$, $e^{\prime}$ and $a^{\prime}$, respectively. Next, we can prove that both actions (eqs.(12) and (20)) have the same `physical' spectrum because we find by using eq.(19) that the two massive scalar bosons have the equal mass
\begin{equation}
m^{2}{\equiv}\frac{e^{2}a(a+1-r^{2})}{a-r^{2}}=\frac{{e^{\prime}}^{2}a^{\prime}(a^{\prime}+1-{r^{\prime}}^{2})}{a^{\prime}-{r^{\prime}}^{2}}{\equiv}{m^{\prime}}^{2}.
\end{equation}
Therefore we have the consequence that the action of the GCSM is self-dual with respect to the generalized (anti-)dualization of $G^{\mu}(\rho)$ and $F_{\mu}(\phi)$.

The self-duality relates in fact to an interchange symmetry of the vector and axial vector current coupling constants of the GCSM. From the fermionic action eq.(11), we introduce $g_{V}$ and $g_{A}$ defined by
\begin{equation}
g_{V}=e, g_{A}=er,
\end{equation}
which are vector and axial vector current coupling constants, respectively. Instead of the parameters {\em e}, {\em r} and {\em a}, we choose $g_{V}$, $g_{A}$ and $m^2$. The transformation eq.(19) can then be expressed as
\begin{eqnarray}
g_{V}&{\longrightarrow}&e^{\prime}={\pm}g_{A},\nonumber \\
g_{A}&{\longrightarrow}&e^{\prime}r^{\prime}={\pm}g_{V},
\end{eqnarray}
and the square of mass $m^2$ is preserved. In accordance with the electric-magnetic duality of the Maxwell theory, we may conclude that the self-duality of the GCSM corresponds to the vector and axial vector current duality. 
\section{Self-duality of the gauge invariant GCSM}
In Ref.[19] two gauge invariant formulations are constructed. One involves a Wess-Zumino term [22] and the other does not. They are equivalent to each other and to the GCSM as well, which means that all of them have the same spectrum. Here we adopt the formulation with the Wess-Zumino term. The complete action reads [19]
\begin{eqnarray}
S=\int d^{2}x\left\{{\frac 1 2}({\partial}_{\mu}{\phi})({\partial}^{\mu}{\phi})+eA^{\mu}({\epsilon}_{{\mu}{\nu}}-rg_{{\mu}{\nu}}){\partial}^{\nu}{\phi}+{\frac 1 2}e^{2}aA_{\mu}A^{\mu}-{\frac 1 4}F_{{\mu}{\nu}}F^{{\mu}{\nu}}\right.\nonumber \\
\left.+{\frac 1 2}(a-r^2)({\partial}_{\mu}{\theta})({\partial}^{\mu}{\theta})+eA^{\mu}\left[r{\epsilon}_{{\mu}{\nu}}+(a-r^2)g_{{\mu}{\nu}}\right]{\partial}^{\nu}{\theta}\right\},
\end{eqnarray}
where ${\theta}(x)$ is known as the Wess-Zumino field. 

In order to investigate the duality with respect to both $\phi$ and $\theta$, we introduce two pairs of auxiliary vector fields $F_{\mu}, G^{\mu}$ and $P_{\mu}, Q^{\mu}$, and replace eq.(24) by the following action
\begin{eqnarray}
S=\int d^{2}x\left\{{\frac 1 2}F_{\mu}F^{\mu}+eA^{\mu}({\epsilon}_{{\mu}{\nu}}-rg_{{\mu}{\nu}})F^{\nu}+{\frac 1 2}e^{2}aA_{\mu}A^{\mu}-{\frac 1 4}F_{{\mu}{\nu}}F^{{\mu}{\nu}}+G^{\mu}(F_{\mu}-{\partial}_{\mu}{\phi})\right.\nonumber \\
\left.+{\frac 1 2}(a-r^2)P_{\mu}P^{\mu}+eA^{\mu}\left[r{\epsilon}_{{\mu}{\nu}}+(a-r^2)g_{{\mu}{\nu}}\right]P^{\nu}+Q^{\mu}(P_{\mu}-{\partial}_{\mu}{\theta})\right\}.
\end{eqnarray}
Variation of eq.(25) with respect to $G^{\mu}$ and $Q^{\mu}$ leads to $F_{\mu}={\partial}_{\mu}{\phi}$ and $P_{\mu}={\partial}_{\mu}{\theta}$, which shows the equivalence between eqs.(24) and (25). In addition, variation of eq.(25) with respect to $F_{\mu}$ and $P_{\mu}$ gives the relations
\begin{eqnarray}
F_{\mu}&=&-G_{\mu}+e({\epsilon}_{{\mu}{\nu}}+rg_{{\mu}{\nu}})A^{\nu},\nonumber \\
P_{\mu}&=&-\frac{1}{a-r^2}Q_{\mu}+e\left(\frac{r}{a-r^2}{\epsilon}_{{\mu}{\nu}}-g_{{\mu}{\nu}}\right)A^{\nu}.
\end{eqnarray} 
Substituting eq.(26) into eq.(25), we obtain the dual action of the gauge invariant GCSM
\begin{eqnarray}
S_{dual}=\int d^{2}x\left[-{\frac 1 2}G_{\mu}G^{\mu}-eA^{\mu}({\epsilon}_{{\mu}{\nu}}-rg_{{\mu}{\nu}})G^{\nu}+\frac{e^{2}a}{2(a-r^2)}A_{\mu}A^{\mu}-{\frac 1 4}F_{{\mu}{\nu}}F^{{\mu}{\nu}}\right.\nonumber \\
\left.-\frac{1}{2(a-r^2)}Q_{\mu}Q^{\mu}-eA^{\mu}\left(\frac{r}{a-r^2}{\epsilon}_{{\mu}{\nu}}+g_{{\mu}{\nu}}\right)Q^{\nu}+{\phi}{\partial}_{\mu}G^{\mu}+{\theta}{\partial}_{\mu}Q^{\mu}\right].
\end{eqnarray}
Varying eq.(27) with respect to both $\phi$ and $\theta$, we obtain the equations ${\partial}_{\mu}G^{\mu}=0$ and ${\partial}_{\mu}Q^{\mu}=0$, whose solution should be
\begin{eqnarray}
G^{\mu}(\rho)={\pm}{\epsilon}^{{\mu}{\nu}}{\partial}_{\nu}{\rho}{\equiv}{\pm}{\epsilon}^{{\mu}{\nu}}F_{\nu}(\rho),\nonumber \\
Q^{\mu}(\eta)={\mp}{\epsilon}^{{\mu}{\nu}}{\partial}_{\nu}{\eta}{\equiv}{\mp}{\epsilon}^{{\mu}{\nu}}P_{\nu}(\eta),
\end{eqnarray}
where ${\rho}(x)$ and ${\eta}(x)$ are arbitrary scalar fields. When eq.(28) is substituted into eq.(27), the dual action is expressed in terms of $\rho$ and $\eta$
\begin{eqnarray}
S_{dual}=\int d^{2}x\left[{\frac 1 2}({\partial}_{\mu}{\rho})({\partial}^{\mu}{\rho}){\pm}eA^{\mu}(r{\epsilon}_{{\mu}{\nu}}-g_{{\mu}{\nu}}){\partial}^{\nu}{\rho}+\frac{e^{2}a}{2(a-r^2)}A_{\mu}A^{\mu}-{\frac 1 4}F_{{\mu}{\nu}}F^{{\mu}{\nu}}\right.\nonumber \\
\left.+\frac{1}{2(a-r^2)}({\partial}_{\mu}{\eta})({\partial}^{\mu}{\eta}){\pm}eA^{\mu}\left({\epsilon}_{{\mu}{\nu}}+\frac{r}{a-r^2}g_{{\mu}{\nu}}\right){\partial}^{\nu}{\eta}\right].
\end{eqnarray} 

In order to compare the action eq.(24) with its dual partner eq.(29), we introduce, as we did in the previous section, three new parameters $r^{\prime}$, $e^{\prime}$ and $a^{\prime}$ defined by
\begin{equation}
r^{\prime}=\frac{1}{r}, e^{\prime}={\pm}er, a^{\prime}=\frac{a}{(a-r^2)r^2},
\end{equation}
where only the third is different from that of eq.(19), and then rewrite eq.(29) as
\begin{eqnarray}
S_{dual}=\int d^{2}x\left\{{\frac 1 2}({\partial}_{\mu}{\rho})({\partial}^{\mu}{\rho})+e^{\prime}A^{\mu}({\epsilon}_{{\mu}{\nu}}-r^{\prime}g_{{\mu}{\nu}}){\partial}^{\nu}{\rho}+{\frac 1 2}{e^{\prime}}^{2}a^{\prime}A_{\mu}A^{\mu}-{\frac 1 4}F_{{\mu}{\nu}}F^{{\mu}{\nu}}\right.\nonumber \\
\left.+{\frac 1 2}(a^{\prime}-{r^{\prime}}^{2})({\partial}_{\mu}{\eta})({\partial}^{\mu}{\eta})+e^{\prime}A^{\mu}\left[r^{\prime}{\epsilon}_{{\mu}{\nu}}+(a^{\prime}-{r^{\prime}}^{2})g_{{\mu}{\nu}}\right]{\partial}^{\nu}{\eta}\right\},
\end{eqnarray} 
which has the same form as eq.(24) with the replacements of $\phi$, $\theta$, {\em r}, {\em e} and {\em a} by $\rho$, $\eta$, $r^{\prime}$, $e^{\prime}$ and $a^{\prime}$, respectively. Particularly, we discover that eq.(21) is still satisfied to the reparameterization (eq.(30)), which shows the equivalence between eqs.(24) and (31). Consequently, the gauge invariant formulation of the GCSM has self-duality with respect to the generalized dualization (anti-dualization) of $G^{\mu}(\rho)$ and $F_{\mu}(\phi)$ and anti-dulaization (dualization) of $Q^{\mu}(\eta)$ and $P_{\mu}(\theta)$.

Note that the transformation eq.(30) can also be expressed as eq.(23) and the square of mass is preserved if we equivalently utilize $g_{V}$, $g_{A}$ and $m^2$. This means, as we stated in the previous section, that the self-duality relates to an interchange symmetry of the vector and axial vector current coupling constants.
\section{Conclusion}
We have shown that the self-duality indeed appears in the interacting model of Floreanini-Jackiw chiral bosons and gauge fields, the generalized chiral Schwinger model (GCSM) and its gauge invariant formulation that relate to chiral $0$-forms (chiral bosons). The first two models are self-dual with respect to the generalized (anti-)dualization of $F_{\mu}(\phi)$ and $G^{\mu}(\rho)$, and the last is self-dual with respect to the generalized dualization (anti-dualization) of $F_{\mu}(\phi)$ and $G^{\mu}(\rho)$ and anti-dualization (dualization) of $P_{\mu}(\theta)$ and $Q^{\mu}(\eta)$. The word `generalized' means that ${\phi}(x)$ (${\theta}(x)$) does not coincide with ${\rho}(x)$ (${\eta}(x)$) up to a constant on the mass shell. This generalization is reasonable because, as we have clarified, the mass shell is not a necessary condition on duality symmetries of actions. In addition, we have pointed out that the self-duality of both the gauge non-invariant and invariant formulations of the GCSM corresponds to the vector and axial vector current coupling constant duality. Finally, it seems that the self-duality may exist in other models that relate to chiral {\em p}-forms in $D>2$ dimensional space-time. This is now under consideration.
\vskip 48pt
\noindent
{\bf Acknowledgments}
\par
Y.-G. Miao and R. Manvelyan are indebted to the Alexander von Humboldt Foundation for financial
support. Y.-G.M is also supported in part by the National Natural Science Foundation
of China under grant No.19705007 and by the Ministry of Education of China
under the special project for scholars returned from abroad.

\end{document}